\documentclass[twocolumn,aps,prl,superscriptaddress,unsortedaddress,showpacs]{revtex4-1}
\usepackage{amsmath}
\usepackage{amssymb}
\usepackage{graphics}
\usepackage{epsfig}

\begin{document}

\title{Nonlinearly-$\mathcal{PT}$-symmetric systems: spontaneous symmetry
breaking and transmission resonances}
\author{Andrey E. Miroshnichenko$^{1}$, Boris A. Malomed$^{2}$, and Yuri S.
Kivshar$^{1}$}
\affiliation{$^{1}$Nonlinear Physics Centre,
Australian National University, Canberra ACT
0200, Australia\\
$^2$Department of Physical Electronics, School of Electrical
Engineering, Faculty of Engineering, Tel Aviv University, Tel Aviv
69978, Israel}

\begin{abstract}
We introduce a class of \emph{$\mathcal{PT}$}-symmetric systems
which include mutually matched $\emph{nonlinear}$ loss and gain (in
other words, a class of \emph{$\mathcal{PT}$}-invariant Hamiltonians
in which both the harmonic and anharmonic parts are non-Hermitian).
For a basic system in the form of a dimer, symmetric and asymmetric
eigenstates, including multistable ones, are found analytically. We
demonstrate that, if coupled to a linear chain, such a nonlinear
$\mathcal{PT}$-symmetric dimer generates new types of nonlinear
resonances, with the completely suppressed or greatly amplified
transmission, as well as a regime similar to the
electromagnetically-induced transparency (EIT). The implementation
of the systems is possible in various media admitting controllable
linear and nonlinear amplification of waves.
\end{abstract}

\maketitle

\textit{Introduction}. In the past few years, the study of systems
exhibiting the parity-time ($\mathcal{PT}$) symmetry has drawn a great deal
of attention. The underlying idea is to extend canonical quantum
mechanics~by introducing a class of non-Hermitian Hamiltonians which may,
nevertheless, exhibit entirely real eigenvalue spectra~\cite%
{Bender:1998-5243:PRL}. A necessary condition for the Hamiltonian to be $%
\mathcal{PT}$-symmetric is that its linear-potential part $V(x)$, being
complex, is subject to the spatial-symmetry constraint, $V(x)=V^{\ast }(-x)$%
. The complex $\mathcal{PT}$-symmetric potentials can be realized in the
most straightforward way in optics, by combining the spatial modulation of
the refractive index with properly placed gain and loss~\cite%
{Ruschhaupt:2005-L171:JPA}. This possibility has stimulated extensive
theoretical~\cite{Berry:2008-244007:JPA} and experimental~\cite%
{Guo:2009-93902:PRL} studies.

In the $\mathcal{PT}$-symmetric Hamiltonians introduced in the context of
the field theory and optics, solely the harmonic part features the matched
gain and loss, while the anharmonic part, if any, is Hermitian, giving rise
to nonlinear dynamical models in which only the linear part features the
balanced dissipation and amplification \cite{nonlinear}. In this work, we
put forward an extension of the $\mathcal{PT}$ symmetry, in the form of
Hamiltonians whose anharmonic part too includes the mutually matched loss
and gain. Solving the corresponding dynamical equations, we demonstrate that
such \emph{nonlinearly}-$\mathcal{PT}$-symmetric systems also give rise to
eigenstates with real frequencies. New findings, which are specific to the
systems with the matched nonlinear gain and loss, are eigenstates with a
spontaneously broken spatial symmetry (that, as mentioned above, implements
the $\mathcal{PT}$ symmetry of the Hamiltonian), and multistability of
eigenstates. A straightforward application of these states is realized by
coupling the $\mathcal{PT}$ system to linear chains: we demonstrate that
this setting gives rise to new types of multistable nonlinear Fano
resonances, as well as to transmission regimes with a very strong
amplification, and those similar to EIT (electromagnetically-induced
transparency). Systems of this type can be implemented in optics, using
saturable absorbers \cite{SESAM} and two-photon losses to realize the
nonlinear $\mathcal{PT}$ symmetry, as well as in any medium which allows
nonlinear amplification of waves, such as cavity polaritons \cite{polariton}%
, surface plasmons \cite{plasmon}, and magnons \cite{magnon}.

We start by introducing a solvable nonlinearly-$\mathcal{PT}$-symmetric
system, in the form of a dimer, in which the symmetric linear gain and loss
terms come along with their nonlinear mutually conjugate counterparts,
\begin{equation}
\begin{array}{l}
{i\dot{\psi}_{A}=(E+i\gamma _{0}-i\gamma _{2}|\psi _{A}|^{2}+\chi |\psi
_{A}|^{2})\psi _{A}+V\psi _{B},} \\*[9pt]
{i\dot{\psi}_{B}=(E-i\gamma _{0}-i\gamma _{2}|\psi _{A}|^{2}+\chi |\psi
_{B}|^{2})\psi _{B}+V\psi _{A}.}%
\end{array}
\label{v1}
\end{equation}%
Here the overdot stands for the time derivative, $\gamma _{0}>0$ accounts
for the linear gain and loss, which act on complex variables $\psi _{A}$ and
$\psi _{B}$, respectively, $E$ is a frequency shift (with respect to a
linear chain if the dimer is coupled to it, see below), $\gamma _{2}$
accounts for the $\mathcal{PT}$-symmetric nonlinear\ loss and gain (as shown
below, stable eigenstates are obtained with $\gamma _{2}>0$, i.e., if the
nonlinear loss competes with the linear gain, and vice versa), $\chi $ is
the strength of the nonlinear frequency shift, and $V$ is a coupling
coefficient.

\textit{Symmetric modes}. Symmetric eigenstates, with $|\psi _{A}|=|\psi
_{B}|$, are sought for as $\psi _{A,B}(t)=A\exp \left( -i\omega t\pm i\delta
/2\right) $, with the amplitude and phase shift determined by the following
equations:%
\begin{equation}
\left[ \chi A^{2}-\left( \omega -E\right) \right] ^{2}+\left( \gamma
_{0}-\gamma _{2}A^{2}\right) ^{2}=V^{2},  \label{A}
\end{equation}%
\begin{equation}
\tan \delta =\left( \gamma _{0}-\gamma _{2}A^{2}\right) \left( \omega
-E-\chi A^{2}\right) ^{-1}~.  \label{tan}
\end{equation}%
Depending on the parameters, Eq. (\ref{A}) may yield no physical solutions
with $A^{2}>0$, a single solution (monostability), and bistability, with two
physical roots. The bistability occurs under conditions%
\begin{equation}
\begin{array}{c}
\left( \omega -E\right) ^{2}>V^{2}-\gamma _{0}^{2}, \\
~\gamma _{0}\gamma _{2}\chi ^{-1}+\left( \omega -E\right) >\sqrt{\left(
\omega -E\right) ^{2}+\gamma _{0}^{2}-V^{2}},%
\end{array}
\label{bi}
\end{equation}%
while the monostability condition is $\left( \omega -E\right)
^{2}<V^{2}-\gamma _{0}^{2}$. Although the system is dissipative, the
symmetric eigenstates with the real frequencies form a continuous family
parameterized by arbitrary frequency $\omega $, which is a manifestation of
the $\mathcal{PT}$ symmetry.

\textit{Asymmetric states and multistability}. The system admits solutions
with broken symmetry too ($A\neq B$): $\psi _{A,B}(t)=\left\{ Ae^{i\delta
/2},Be^{-i\delta /2}\right\} e^{-i\omega t}$, with $\delta $ determined by
the same equation (\ref{tan}) as above. Unlike the symmetric eigenstates,
the asymmetric ones exist at a\emph{\ }single frequency, which is typical to
dissipative systems:%
\begin{equation}
\omega _{\mathrm{AS}}=E+\left( \gamma _{0}/\gamma _{2}\right) \chi .
\label{omega}
\end{equation}%
Further, amplitude $A$ is determined by equation
\begin{equation}
\left( A^{2}\right) ^{2}-\left( \gamma _{0}/\gamma _{2}\right)
A^{2}+V^{2}\left( \chi ^{2}+\gamma _{2}^{2}\right) ^{-1}=0,  \label{V}
\end{equation}%
cf. Eq. (\ref{A}), while the other amplitude is given by $B^{2}=\left(
\gamma _{0}/\gamma _{2}\right) -A^{2},$ which shows that the asymmetric
eigenmode exists only for $\gamma _{2}>0$, i.e., when the nonlinear $%
\mathcal{PT}$-symmetric loss/gain terms compete with their linear
counterparts. We stress that the asymmetric solutions are supported by the
balance of the nonlinear gain and loss, as they do not exist for $\gamma
_{2}=0$.

The above relations yield two physical solutions (i.e., the bistability) for
the asymmetric modes, with $A^{2},B^{2}>0$, at $\chi ^{2}/\gamma
_{2}^{2}>4\left( V^{2}/\gamma _{0}^{2}\right) -1$, and no solutions in the
opposite case. Under this condition, inequalities (\ref{bi}) hold too for $%
\omega =\omega _{\mathrm{AS}}$, i.e., the system gives rise to the \textit{%
multistability}, with four coexisting eigenstates, two symmetric and two
asymmetric.

\begin{figure}[tbp]
\includegraphics[width=.7\columnwidth]{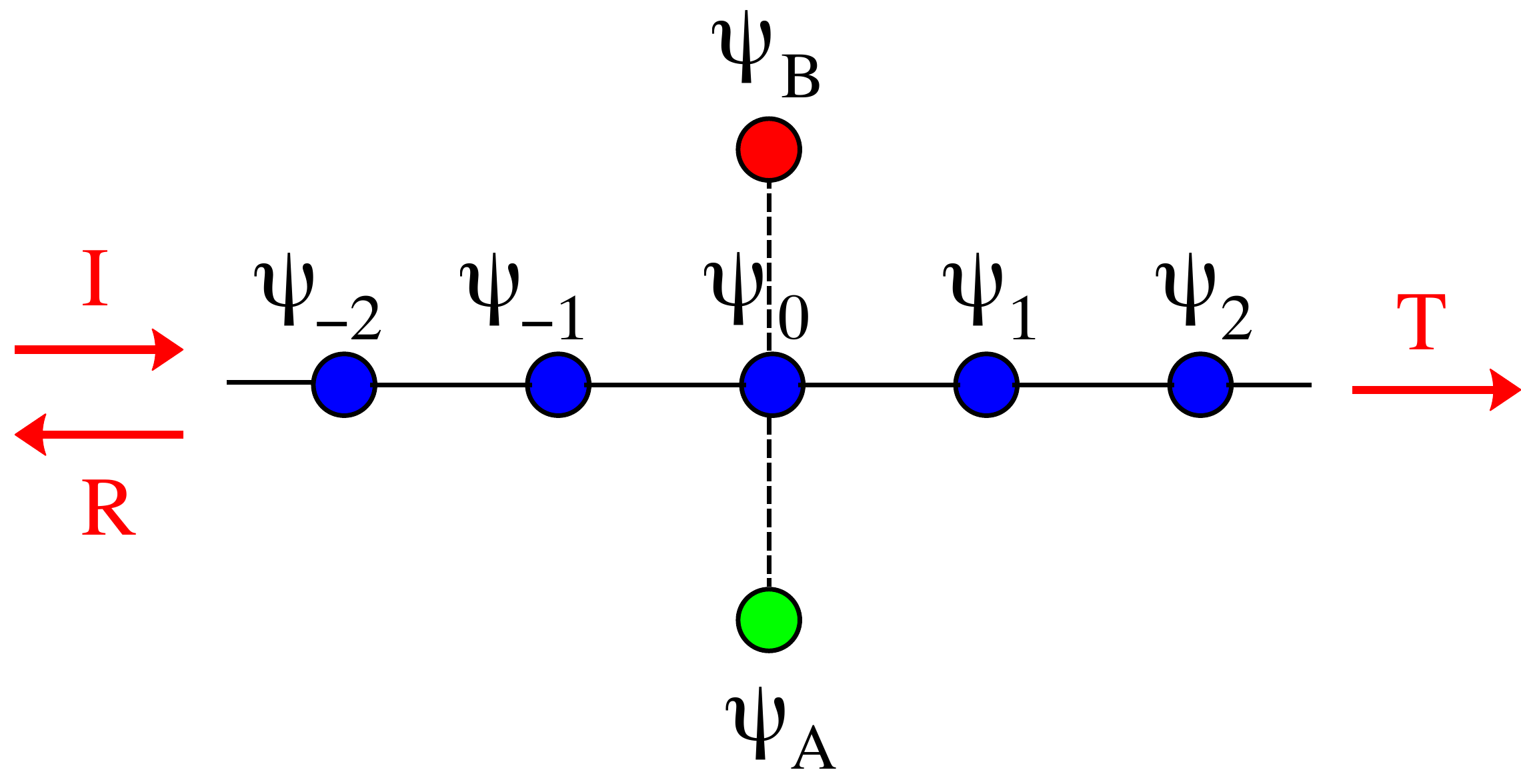}
\caption{(Color online) The linear chain with the side-coupled elements
featuring the nonlinear $\mathcal{PT}$ \ symmetry. The arrows indicate
incident, reflected and transmitted waves.}
\label{fig:fig1}
\end{figure}

\textit{The scattering problem}. The next step is to couple the dimer to a
chain transmitting linear discrete waves $\psi _{n}(t)$, as shown in Fig.~%
\ref{fig:fig1} (note that the entire system remains $\mathcal{PT}$%
-symmetric). Here we focus on the most fundamental version of the system,
with $\chi =0$, the nonlinearity being represented by the matched cubic loss
and gain, the respective coupled system being
\begin{eqnarray}
i\dot{\psi}_{A} &=&E\psi _{A}+i\left( \gamma _{0}-\gamma _{2}|\psi
_{A}|^{2}\right) \psi _{A}+V\psi _{0},  \label{psiA} \\
i\dot{\psi}_{n} &=&C\left( \psi _{n-1}+\psi _{n+1}\right) +V\delta
_{n,0}\left( \psi _{A}+\psi _{B}\right) ,  \label{psi0} \\
i\dot{\psi}_{B} &=&E\psi _{B}-i\left( \gamma _{0}-\gamma _{2}|\psi
_{B}|^{2}\right) \psi _{B}+V\psi _{0},  \label{psiB}
\end{eqnarray}%
where $C$ is the coupling constant in the linear chain.
The general solution corresponding to the scattering of incident waves with
amplitude $I$ on the $\mathcal{PT}$ complex is looked for as%
\begin{equation}
\psi _{n}=\left\{
\begin{array}{c}
Ie^{i(kn-\omega t)}+Re^{-i(kn+\omega t)}\ \left( n\leq 0\right) , \\
Te^{i(kn-\omega t)}~\left( n\geq 0\right) ,%
\end{array}%
\right.   \label{incident}
\end{equation}%
where wavenumber $k>0$ is determined by the dispersion equation for the
linear chain, $k=\cos ^{-1}\left( \omega /2C\right) $, while $R$ and $T$ are
the amplitudes of reflected and transmitted waves. A straightforward
analysis of Eqs. (\ref{psi0},\ref{incident}) at $n=0$ yields $R=\psi _{0}-I$%
, $T=\psi _{0}$, and the expression for $\psi _{0}$ in terms $I$ and $\psi
_{A,B}^{(0)}$:
\begin{equation}
\psi _{0}=I+iV\left( 2C\sin k\right) ^{-1}\left( \psi _{A}^{(0)}+\psi
_{B}^{(0)}\right) .  \label{Psi0}
\end{equation}%
The substitution of expression (\ref{Psi0}) into the stationary version of
Eqs. (\ref{psiA}) and (\ref{psiB}) leads to a system of complex cubic
equations:

\begin{gather}
\left( E-\omega \right) \psi _{A,B}^{(0)}+iV^{2}\left( 2C\sin k\right)
^{-1}\left( \psi _{A}^{(0)}+\psi _{B}^{(0)}\right)  \notag \\
\pm i\left( \gamma _{0}-\gamma _{2}\left\vert \psi _{A,B}^{(0)}\right\vert
^{2}\right) \psi _{A,B}^{(0)}=-VI.  \label{JB}
\end{gather}%
One should solve Eq. (\ref{JB}) for $\psi _{A,B}^{(0)}$ at given $I$ and $%
\omega $. Then, $\psi _{0}$ can be found from Eq. (\ref{Psi0}), and,
eventually, the reflection and transmission coefficients can be found.

\begin{figure}[tbp]
\includegraphics[width=.7\columnwidth]{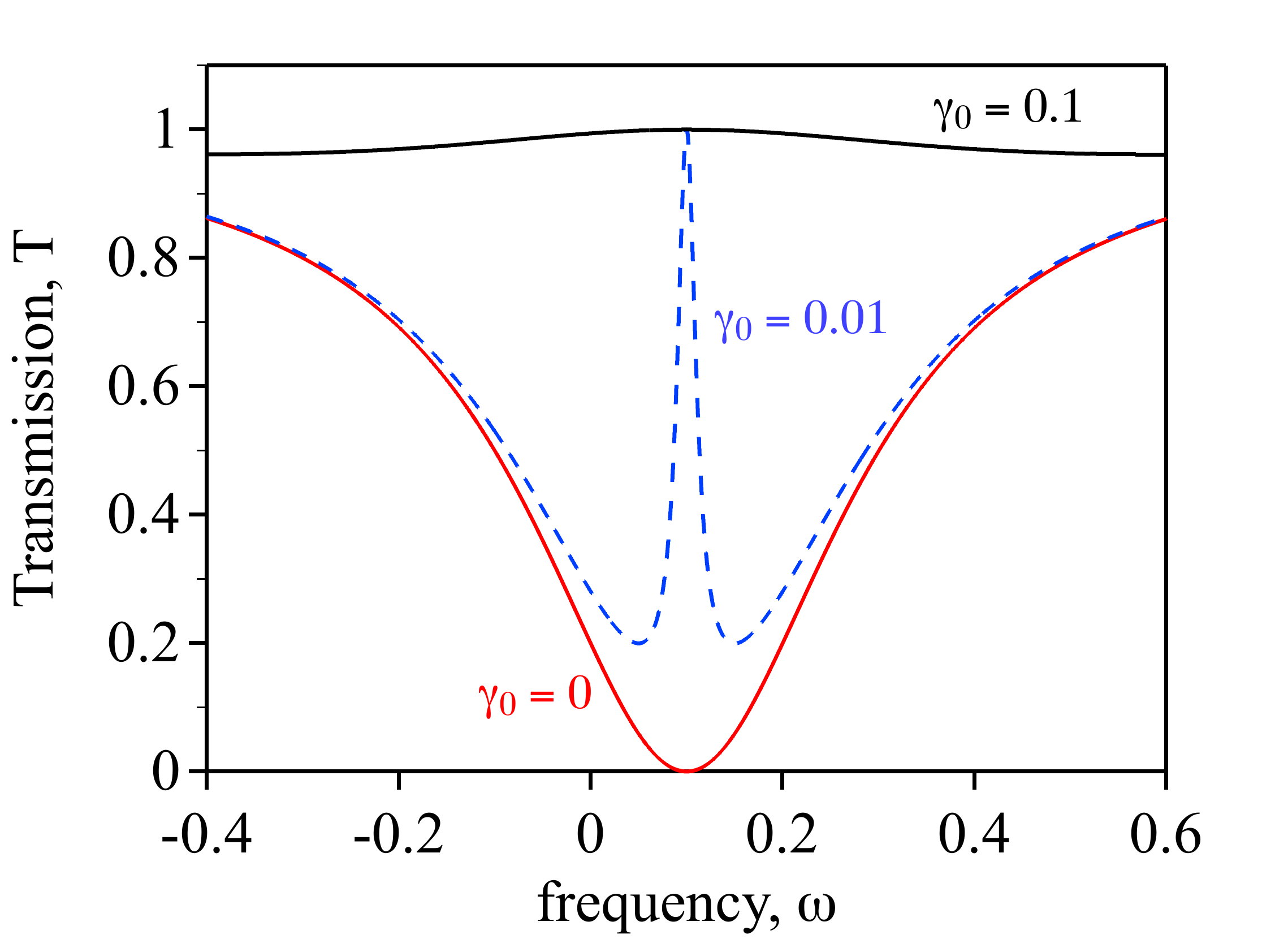}
\caption{(Color online) The transmission coefficient for several values of
the gain/loss factor $\protect\gamma _{0}$ in the linear system ($\protect%
\gamma _{2}=0$). Other parameters are $E=0.1$, $V=0.2$, and $C=1$. }
\label{fig:fig2}
\end{figure}

\textit{The scattering in the symmetric regime}. In the linear system ($%
\gamma _{2}=0$), Eq. (\ref{JB}) yields only symmetric solutions, with $|\psi
_{A}|=|\psi _{B}|$. The corresponding scattering spectrum, displayed in Fig.~%
\ref{fig:fig2}, demonstrates two noteworthy effects. One is the
suppression of the transmission by the degenerate side-coupled
elements without the gain and loss, $\gamma _{0}=0$. In this case,
the eigenfrequencies of both elements are identical, and their
excitation results in the resonant reflection at $\omega =E$, which
can be explained in terms of the Fano
resonance~\cite{rmp}. The presence of the weak linear gain and loss, with $%
\gamma _{0}\ll 1$, lifts the degeneracy between the attached sites, leading
to a response resembling the electromagnetically-induced transparency (EIT)
\cite{EIT}, with the total transmissivity ($T=1$) at $\omega =E$, between
resonant reflections on the pair of slightly detuned linear $\mathcal{PT}$
elements.

\begin{figure}[tbp]
\includegraphics[width=.7\columnwidth]{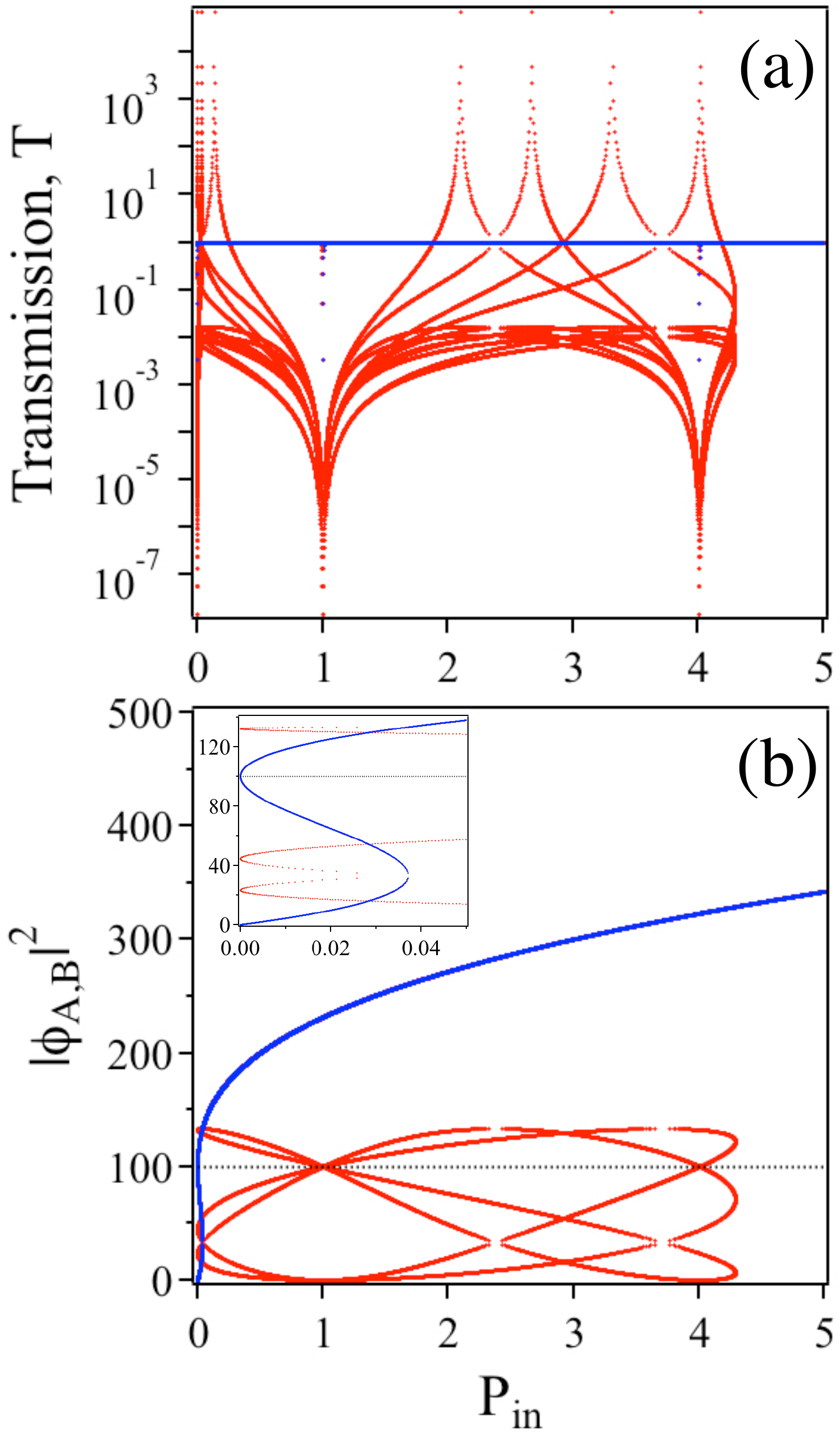}
\caption{(Color online) The transmission coefficient (a) and excitation
intensity of the side-coupled $\mathcal{PT}$ elements (b) for $\protect%
\gamma _{0}=0.01$, $\protect\gamma _{2}=0.0001$, $V=0.2$, $C=1$, and $%
\protect\omega =E=0.1$. The nonlinear Fano resonances correspond to
$T=0$. The red and blue curves depict, respectively, the full
multitude of asymmetric scattering regimes, produced by Eq.
(\protect\ref{JB}), and the symmetric one corresponding to Eq.
(\protect\ref{cubic}). The inset in (b) zooms the tristability
region in the latter case. The horizontal dotted line corresponds to
Eq. (\protect\ref{delta}).} \label{fig:fig3}
\end{figure}

In the system combining the linear chain and the nonlinear $\mathcal{PT}$
scatterer with $\gamma _{2}>0$, one can find symmetric solutions with $\psi
_{A}^{(0)}=-\psi _{B}^{(0)}=-i\phi $, where $\phi $ is real. First, we
consider the case of $\phi \neq \sqrt{\gamma _{0}/\gamma _{2}}$ (this value
plays a special value, as shown below). Then, the symmetric mode exists at $%
\omega =E$, with $\phi $ determined by equation
\begin{equation}
\gamma _{2}\phi ^{3}-\gamma _{0}\phi +VI=0,  \label{cubic}
\end{equation}
which yields a single real root for $P_{\mathrm{in}}\equiv I^{2}>\left(
4/27\right) \gamma _{0}^{3}/\left( V^{2}\gamma _{2}\right) $, and three real
solutions (\textit{tristability}) in the opposite case. According to Eq. (%
\ref{Psi0}), all these solutions realize the perfect EIT-like
transmissivity, with $T\equiv 1$ [the horizontal blue line in Fig. \ref%
{fig:fig3}(a)]. The family of the symmetric states is displayed by the blue
curves in Fig.~\ref{fig:fig3}(b), where the tristability occurs at $P_{%
\mathrm{in}}<1/27$.

\textit{Nonlinear Fano resonances}. In contrast to its linear counterpart,
the nonlinear system may support complete suppression of the transmission ($%
T=0$), i.e., nonlinear Fano resonances~\cite{rmp}. From Eq. (\ref{Psi0}) it
follows that $\psi _{A}^{(0)}+\psi _{B}^{(0)}=2iICV^{-1}\sin k$ for $\psi
_{0}=T=0$. The substitution of this into Eq. (\ref{JB}) leads to the system%
\begin{equation}
\left( E-\omega \right) \psi _{A,B}^{(0)}\pm i\left( \gamma _{0}-\gamma
_{2}\left\vert \psi _{A,B}^{(0)}\right\vert ^{2}\right) \psi _{A,B}^{(0)}=0,
\label{simpleB}
\end{equation}%
which gives rise to a continuous family of symmetric nonlinear Fano
resonances, with $\omega =E$ and
\begin{equation}
\psi _{A,B}^{(0)}=i\sqrt{\gamma _{0}/\gamma _{2}}\exp \left( \pm i\delta
\right) ,  \label{delta}
\end{equation}%
\begin{equation}
\cos \delta =\left( 2V\right) ^{-1}\sqrt{\left( 4C^{2}-E^{2}\right) \left(
\gamma _{2}/\gamma _{0}\right) P_{\mathrm{in}}}  \label{final}
\end{equation}%
(recall the above EIT-like symmetric family, corresponding to $T\equiv 1$,
had $\phi \neq \sqrt{\gamma _{0}/\gamma _{2}}$ and $\cos \delta =0$).
Equation (\ref{final}) imposes condition $\cos ^{2}\delta \leq 1$, i.e.,
\begin{equation}
P_{\mathrm{in}}\leq 4V^{2}\left( \gamma _{0}/\gamma _{2}\right) \left(
4C^{2}-E^{2}\right) ^{-1}.  \label{P_max}
\end{equation}
Thus, at $\omega =E$, the continuous family of the symmetric nonlinear Fano
resonances exists in this interval of the intensity of the incident wave.
The novelty of the result is that the Fano resonance is usually obtained as
an isolated solution.

\textit{The asymmetric scattering regimes}. Equation (\ref{simpleB}) for the
nonlinear Fano resonances also admits two \textit{ultimate asymmetric states}%
, with the vanishing excitation at one of the $\mathcal{PT}$ elements: $%
\omega =E$ and
\begin{equation}
\psi _{A}^{(0)}=\sqrt{\gamma _{0}/\gamma _{2}},\psi _{B}^{(0)}=0,
\label{ultim}
\end{equation}%
or vice versa, with $A\rightleftarrows B$. In either case, this solution
exists at $P_{\mathrm{in}}=V^{2}\left( \gamma _{0}/\gamma _{2}\right) \left(
4C^{2}-E^{2}\right) ^{-1}$. This point falls into the range (\ref{P_max}) of
the existence of the symmetric Fano-resonance solutions, which implies
intrinsic bistability of the nonlinear Fano resonances.

Equation (\ref{JB}) gives rise to other asymmetric scattering regimes,
which, in particular, may produce a strong resonant amplification of the
transmitted wave. The complete set of the asymmetric scattering states is
depicted by the red curves in Fig.~\ref{fig:fig3}.

\begin{figure}[tbp]
\includegraphics[width=.9\columnwidth]{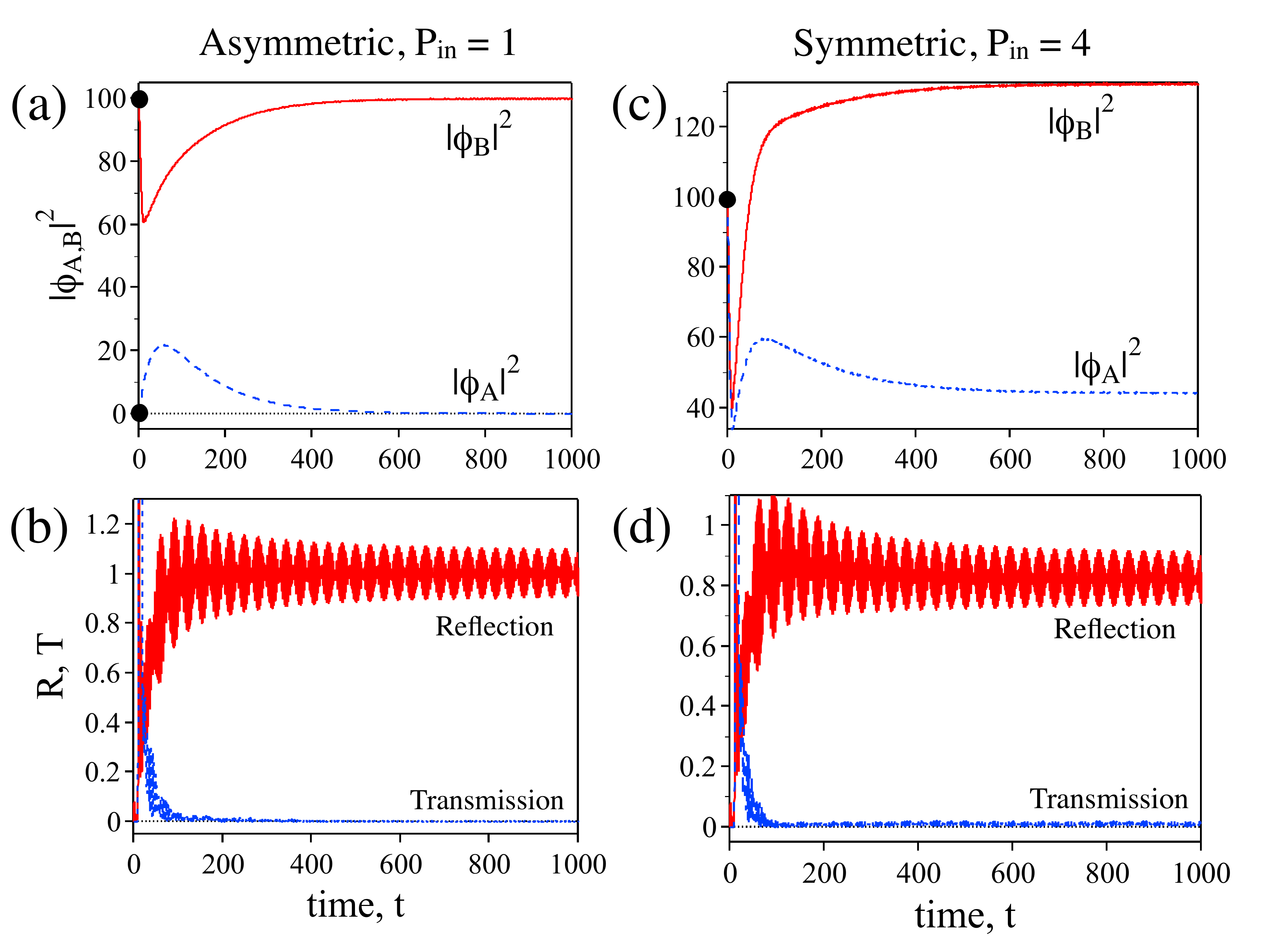}
\caption{(Color online) The perturbed evolution of those asymmetric (left)
and symmetric (right) Fano-resonance modes from Fig.~\protect\ref{fig:fig3}
which are dynamically unstable. Filled circles indicate the initial states.}
\label{fig:fig4}
\end{figure}

\textit{Stability}. The stability of the above analytical solutions was
checked in direct simulations of Eqs. (\ref{psiA})-(\ref{psiB}). The results
demonstrate that the ultimate asymmetric state (\ref{ultim}) with the
nonzero excitation at the linear-loss element is stable, while its
counterpart with the excitation at the linear-gain element is unstable,
transforming itself into an oscillatory mode (apparently, a limit cycle),
see left panels in Fig.~\ref{fig:fig4}. These results can be readily
understood following the similarity to previously studied systems composed
of coupled cores with the linear gain and loss acting separately in them,
which also give rise to a pair of stable and unstable modes \cite{Winful}.

Symmetric Fano-resonance modes (\ref{delta}) are unstable, also developing
intrinsic oscillations, with a very low transmissivity and spontaneously
broken symmetry between the $\mathcal{PT}$ elements, $|\psi _{A}|\neq |\psi
_{B}|$, as shown in the right panels in Fig.~\ref{fig:fig4}. In fact, such
asymmetric states correspond to the nearly perfect Fano resonance in Fig.~%
\ref{fig:fig3} at $P_{\mathrm{in}}=4$. The resonantly-amplified
transmission regimes are also unstable, due to enhancement of the
field on the linear-gain element.

\textit{Nonpropagating modes}. The dispersion relation for Eq. (\ref{psi0})
demonstrates that frequencies of the propagating waves belong to the
respective phonon band, $|\omega |<2C$. Applying an excitation above the
band, with $\omega >2C$, it is possible to find the corresponding exact
solutions for localized modes pinned to the $\mathcal{PT}$ complex:%
\begin{equation}
\psi _{n}=\frac{V\left( \tilde{\psi}_{A}+\tilde{\psi}_{B}\right) }{2\omega -%
\sqrt{\omega ^{2}+4C^{2}}}\left( \frac{\sqrt{\omega ^{2}+4C^{2}}+\omega }{2C}%
\right) ^{-|n|},  \label{nonpro}
\end{equation}%
where $\tilde{\psi}_{A}$ and $\tilde{\psi}_{B}$ are given by the above
solutions for the symmetric and asymmetric modes, i.e., severally, Eqs. (\ref%
{A}), (\ref{tan}) and (\ref{omega}), (\ref{V}), with $V$ and $E$ replaced by
$\tilde{V}\equiv V^{2}/\left( 2\omega -\sqrt{\omega ^{2}+4C^{2}}\right) $
and $\tilde{E}\equiv E+\tilde{V}$. This means that the solution for the
nonpropagating symmetric modes remains explicit, while Eq. (\ref{omega}) for
the asymmetric mode takes the form of a quartic equation for $\omega _{%
\mathrm{AS}}$.

\textit{In conclusion}, we have introduced the species of $\mathcal{PT}$%
-symmetric systems with the balanced nonlinear gain and loss. For the dimer
system, we have produced a complete set of analytical solutions, which
feature the spontaneous symmetry breaking and multistability, not achievable
in previously studied $\mathcal{PT}$-symmetric systems. We have
demonstrated, also in the analytical form, that the symmetric and asymmetric
excitations in the dimer, if it is coupled to the linear chain, give rise to
a variety of novel nonlinear Fano resonances, including the bistability
between them, and simultaneously to perfect-transmission regimes resembling
EIT, as well as to the resonantly-amplified transmission. The coexistence of
these scattering channels suggests an application to the design of
data-processing schemes. Nonpropagating modes in the chain, pinned to the $%
\mathcal{PT}$ scatterer, were found too.

B.A.M. thanks Nonlinear Physics Centre at the Australian National University
and Department of Telecommunications at the University of New South Wales
for their hospitality.

\end{document}